\documentclass[11pt,a4article,final]{article}
\usepackage[top=2.5cm,bottom=1.5cm,left=2.4cm,right=1.5cm]{geometry}
\usepackage{lipsum}

\usepackage[utf8]{inputenc}
\usepackage[T1]{fontenc}
\usepackage[english]{babel}
\usepackage{amsmath}
\usepackage{multirow}
\usepackage{amsfonts}
\usepackage{amssymb}
\usepackage{makeidx}
\usepackage{natbib}
\usepackage{tfrupee}
\usepackage{graphicx}
\usepackage{ifvtex}
\usepackage{setspace}
\usepackage{tabu}
\usepackage{booktabs}
\usepackage{longtable}
\usepackage{rotfloat}
\usepackage{tikz}
\usepackage{color,hyperref}
\usetikzlibrary{backgrounds}
\usetikzlibrary{arrows}
\usetikzlibrary{shapes,shapes.geometric,shapes.misc}
\pgfdeclarelayer{edgelayer}
\pgfdeclarelayer{nodelayer}
\pgfsetlayers{background,edgelayer,nodelayer, main}
\tikzstyle{none}=[inner sep=0mm]
\tikzset{new style 0/.style={circle,draw}}

\newtheorem{proposition}{Proposition}

\newtheorem{definition}{Definition}
\newenvironment{proof}{\paragraph{Proof:}}{\hfill$\square$}
\title{Demand Analysis and Customized Product Offering Design on E-Commerce Platform}
\author{Dipankar Das\\Assistant Professor, Goa Institute of Management\\Email ID:dipankar3das@gmail.com;dipankar@gim.ac.in}
\date{}
\begin{document}
	\maketitle
	
	\begin{abstract}
It can be observed that the purchasing decision of an individual consumer in an electronic marketplace is determined by a set of factors, such as personal characteristics of the consumer, product pricing, minimum price-quantity combination offered, decision-making space, and underlying motivation of the consumer. These factors are combined to form a consumer's choice problem domain, which plays a pivotal role in the product offering. In this study, we attempt to focus on how the products? Offered can be customized by incorporating the quantity and pack size of the products along with the factors above to form a more extensive domain for examining the combined effects of all of these factors on demand. Accordingly, the demand function is defined by a novel method invoking the extended domain of choice problem in the electronic marketplace. Consequently, the predictable uncertainty associated with the consumer's demand function may disappear, increase the likelihood of earning optimum revenue through customized combinations of the components of the extended domain of choice problem, and improve the understanding of the fluctuations in consumer demand. Finally, we propose a generalized price response function with standard properties applicable to E-Commerce. 

	\end{abstract}
	\section{Introduction}
\begin{figure}[H]
	\centering\includegraphics[scale=0.20]{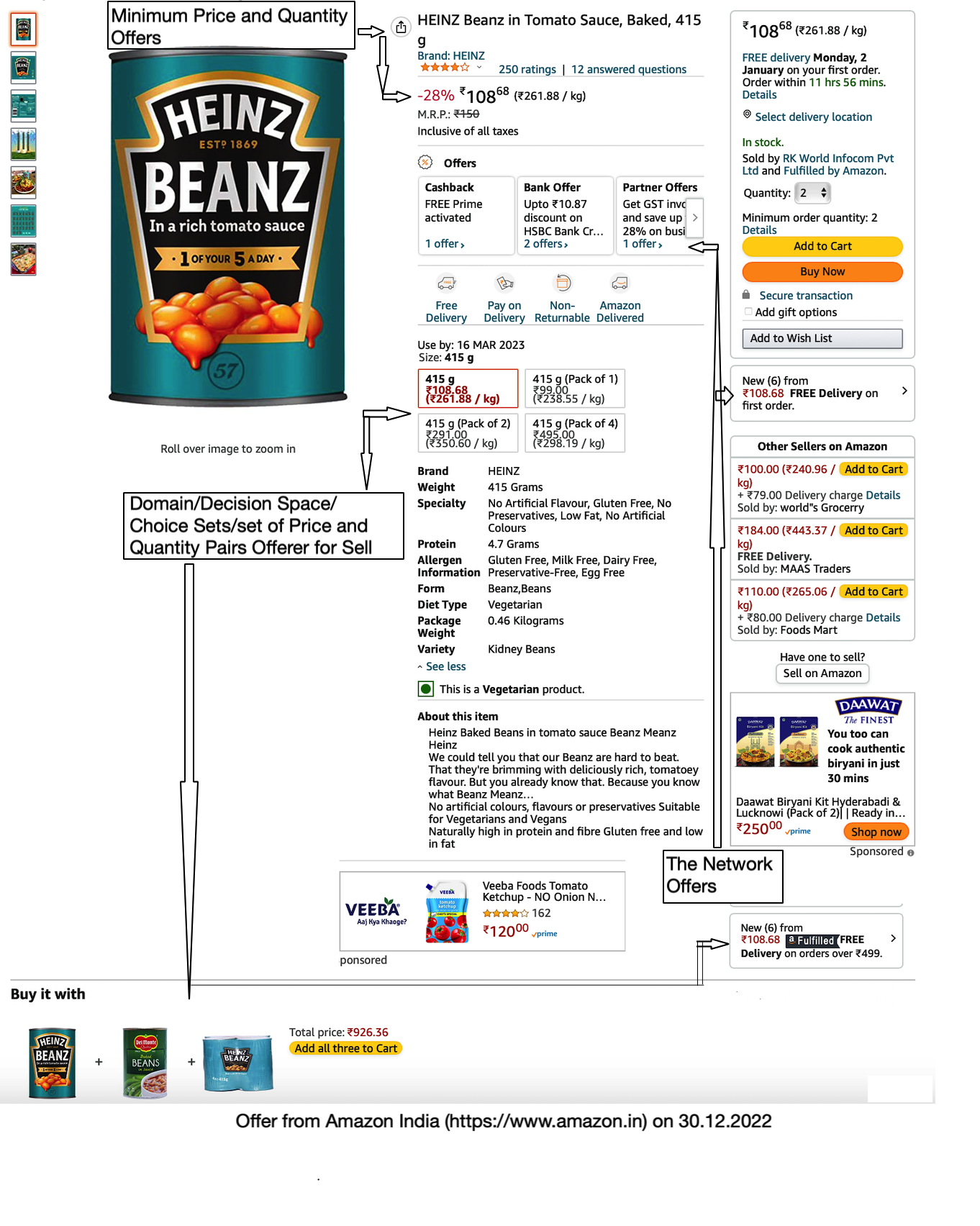}\caption{Pricing of a product by Amazon India.}\label{figure30}
\end{figure}

	It is rather common in the E-commerce marketplace that the decision to purchase an item/product by any customer depends on the customer's characteristics, the price of the product, offered minimum price-quantity combinations, decision-making space, and intention of the customer. These factors are combined to form a proposed option set for the customers (which may be considered a domain or decision set). From this domain a customer chooses the most preferred bundle/set ((refer to Figure \ref{figure30})\footnote{In Figure \ref{figure30}; \rupee = INR. $ 1 $ USD $ \approx 83.04 $ INR Feb 11 at 10:03 PM GMT+5:30 .}). Based on the merged combination of customers request and its revenue/profit margin, a particular platform creates a choice set for the buyers and displays it for selection. \\ 
	\indent Supposing a potential customer desires to buy a product from a well-known and widely used e-commerce platform, Amazon, that customer starts searching for the desired products by setting some filtering criteria. In Amazon's web browser, these criteria are sequentially organized as follows: prime new stores, department, customer ratings, brand, price range (maximum and minimum), deals, category of items, membership service, specialty, cuisine, discount, availability, etc. When purchasing an item/product, any customer selects the best option for every criterion and then waits for the platform's reaction by returning the results. The respective platform reads the selected options based on different criteria and then provides customers with a long list of the items/products that meet their requirements. \\
	\indent Any deviation of choice set offered by the platform from the choice set selected by the customer \textit{de facto} leads to problems in the identification of the individuals as the customers of the products (\cite{rusmevichientong2006nonparametric,aggarwal2004algorithms,calvete2019rank,dominguez2021rank}). Therefore, the aggregate customer demand becomes indeterminate, since this mismatch leads to the shifting of the customers to a different domain where they can accomplish their goal. Under this circumstance, it is hard for any e-commerce platform to forecast the outcome without determining the degree of individual and aggregate utility.\\ 
	\indent Through the earlier mentioned mismatch, any platform can earn supplementary revenue\cite{dominguez2021rank}. Therefore, a platform has two fundamental objectives: to earn revenue due to this mismatch and to retain the customer. To accomplish these objectives, it is necessary to comprehend the customer's demand function. The platform could prepare customized offers by manipulating the domain and knowing the demand function. Usually, a platform chooses a strategy excluding the domain of the consumer, and based on the rationality of the customer, unfavorable outcomes in any platform shift them to a different platform. Each platform caters to different consumer incentives based on their purchase history. As a result, this generates a specific reference point. Accordingly, different platforms meet the objectives of different customers. To understand the mapping of choice from one domain to another, it is essential to perceive the various types of domain/decision spaces, the respective platform's choice set, and its mapping behavior. This understanding enables a platform to analyze and predict the customer's preference pattern and set profit maximizing and envy-free prices\cite{fernandes2016envy,anshelevich2017envy}. Furthermore, it is important here to say that price is one of the components of revenue, and by using this instrument, any firm can maximize its revenue through the balance in supply and demand appropriately\cite{roth2020multidimensional}.\\
	\indent Against this backdrop, there are twofold objectives of this study. Firstly, this study attempts to identify the determinants of the customers' demand for e-commerce platforms. Accordingly, we offer a novel method for determining the demand function of a single sale posting on an e-commerce platform, incorporating many customer characteristics. This will enable the e-commerce platforms to maximize their objectives and satisfy customers. This research describes the topological behavior of the decision set/domain during the first phase. It proposes a new method for calculating the demand function, which depends on the abovementioned factors. In the second phase, the study outlines the customer's criteria for switching from one domain to another and the revenue maximization strategy. The proposed method has been extended from the previous work explained in the revenue-maximizing strategy\cite{das2022understanding,das2022t}.\\
	\indent The second objective is to propose some solutions for the decision-makers, which develop a trade-off between the maximization of sales revenue and retainment of the customers for a particular platform. For the analysis of this study, we mainly focus on the commodities that an e-commerce platform can customize in terms of quantity and pack size. For instance, FMCG products are fruits, vegetables, sugar, dry fruits, cereals, grains, and wheat. \\
	\indent To the best of our knowledge, none of the existing studies proposes a domain's complex topological behavior and a novel approach for determining the customer's demand to avoid the mismatch problem. Thus, this study's expected findings will be important for e-commerce platforms in managing their profit margins and satisfying their customers. Moreover, creating customized offers by anticipating demand is any platform's primary source of additional revenue, reducing fluctuations. We demonstrate that if a consumer's motivation is known, the information presented on an offer may be utilized to ascertain the consumer's perceived demand. Consequently, the proposed strategy can aid in selecting the optimal domain pair and the lowest offered price-quantity pair while maintaining the exogeneity of motives and network offers. This may prevent consumers from switching to a different e-commerce platform.\\
	Moreover, a study shows that online retailers exhibit higher price dispersion than their offline counterparts\cite{aparicio2023pricing}. 
	
	 Answering these questions will allow the platform to analyze and predict the agent/preference consumer's pattern and set profit-maximizing, envy-free prices (\cite{fernandes2016envy}). 
\indent The rest of this paper is structured as follows. In section two, we discuss related literature. The third section describes the model that offers a novel approach to determine the customer demand to overcome the mismatch problem. In section four, we explain the price response function and its properties. Section five discusses the forms and properties of the revenue function. Section six concludes, and section seven discusses the limitations and future scope of the current work. 
\section{Related Literature}
Recently, revenue management (RM) has emerged as one of the most successful areas of operations research applications. The decisions regarding demand management and the processes and systems required to make such decisions are the primary emphasis of RM. The objective is to maximize revenue by managing supply and demand through price fluctuations. According to \cite{talluri2004theory}, the more customer heterogeneity in a supply chain, the greater the possibility to exploit this heterogeneity strategically and tactically to maximize profits. \\
\indent The Price of every item/product in a multi-product pricing model is determined by its position in the list of items\cite{roth2020multidimensional}.
The output of any platform might not reflect the domain of consumers preference. This leads to unsatisfactory outcomes \cite{rusmevichientong2006nonparametric,aggarwal2004algorithms,calvete2019rank,dominguez2021rank}. The finding of another study reveals that online retailers exhibit higher price dispersion than that of their offline counterparts \cite{aparicio2023pricing}.\\
\indent The e-commerce platform already occupies a significant market share despite consumers not minimizing their overall spending on online purchases. This reveals the substantial preference of the consumers in buying from the platform even though they fail to reduce their expenditure from these purchases \cite{das2022non}\\
\indent In addition, greater variation in demand over time due to seasonality, disruptions, or other variables leads to an increase in the difficulty of managing the demand decisions. Consequently, the most essential element of RM is forecasting demand in digital or online markets. According to Foster and Vohra \cite{foster1999regret}, decisions on the risks, benefits, and consequences that are the outcome of the actions initiated by the decision-maker must be taken before the actual course of events in any organization. However, in most e-commerce platforms (as seen in our example), the decision maker's objective may need to be more well-defined, resulting in a matching problem. \\
In the online decision problem, the decision must be made before knowing which state of the world will be obtained. In that case, the loss incurred in each period does not depend on decisions taken in earlier periods. This Regret in the On-line Decision Problem has been studied in Foster and Vohra \cite{FOSTER19997}. The overall goal of pricing is to maximize the discounted profit over a finite sequence of pricing periods. \\
\indent The selection of a profitable price requires knowledge of the demand curve. This knowledge can only be obtained by observation of the demand at different prices in different periods. A model stating that the price variations observed in a market can, in part, be explained by rational learning behavior by firms (\cite{lobo2003pricing}). A revenue management problem has been addressed where the decision is to set the dynamic pricing without knowing the demand function over a finite sales horizon to maximize expected revenues, given an initial inventory (\cite{besbes2009dynamic,farias2010dynamic}.\\
\indent Consumer preference heterogeneity and search frictions have been studied by Morozov et al. \cite{8caa939710e64593a5ecb8a6ea30141e}. Besides traditional methods like interviews and focus groups to identify customer needs for marketing strategy and product development, user-generated content is a promising alternative for identifying customer needs (\cite{timoshenko2019identifying}). \\
\indent Dzyabura and Hauser (\cite{dzyabura2019recommending} propose the product recommendation using consumers preference weights learning has been proposed. In the specific context of revenue management, previous studies should have considered certain assumptions regarding the unpredictability of demand in a marketplace such as an e-commerce platform. As a direct consequence, following the same pricing mechanism strategy may not maximize overall revenue when there is a mismatch due to consumer and seller domain preferences. Our research addresses this gap by examining a set of consumer parameters that may be utilized to accurately derive the demand function and modify the price depending on the anticipated value of the suggested demand function. In particular, our paper proposes that revenue may be maximized to varying degrees by making the domain non-convex. This approach allows managers to arrive at a strategic decision while reducing the amount of ambiguity associated with the consumer demand function.
\section{Model}
The model section has been divided into the proposed demand function and the price response function with properties.  Next, we introduce the following definitions.
\begin{definition}
	Minimum Price-Quantity Pair Offered: On the digital market, the minimum quantity or purchase value of a commodity bundle must be ordered to get the commodity with or without discount offers\footnote{In our example, the "Minimum Price-Quantity Pair Offered" is \rupee 108.68 for 415g}. Customers are not permitted to place orders lower than this pair.
\end{definition}
\begin{definition}
	The domain is the decision space on which the agent makes decisions.
\end{definition}
\begin{definition}
	Decision Space: This is the platform's offering of (price, quantity) pairs. This set has the properties of being both convex and non-convex. The domain will be convex if the commodity prices are linear and non-convex if the commodity prices are non-linear\footnote{In our example, this set is (\rupee 108.68, 415g), (\rupee 99.00, 415g (Pack of 1)), (\rupee 291.00, 415g (Pack of 2)),(\rupee 495.00, 415g (Pack of 4)) and a combo offer of three commodities together at \rupee 926.36. As a result, this offer set is non-convex.}
\end{definition}
\begin{definition}
	Consumer Motive: This is a consumer's revealed preference when comparing two offers or commodity alternatives. In a pair-wise comparison, this could be the preferred option. Each consumer belongs to a market segment and is defined by a set of partial preferences, represented by a directed acyclic graph (DAG) with products as nodes. This is also used in another context in this article. In general, it is a measure of the preference (or degree of importance) of an item in a previous transaction compared to others.
\end{definition}
\subsection{Derivation of the Demand Function and Analysis}
Let a platform prepare and offer different types of choice sets/domains before a customer fulfills the above three definitions viz. (i)\textit{Minimum Price-Quantity Pair Offered}, (ii)\textit{Domain/Decision-space/Set of all Price-Quantity Pairs Offered for Sale}, and (iii)\textit{Consumer Motive}. And under each criterion, a set of finite possibilities is there. Hence, the offer may be prepared and posted by altering the above three criteria. Let the customer want to buy two commodities at a time $ X_{1}\& X_{2} $, i.e., $ X_{i};i=1,2 $. Each commodity can be offered in a finite number of ways called choice sets. Combining these two commodities, three possible domains could be offered. These are (i)\textit{Convex Domain Case with Linear prices of both $ X_{1} $ and $ X_{2} $} ,(ii) \textit{Convex Domain Case with Linear $ X_{1} $ Price and Non-Linear $ X_{2} $ Price},(iii)\textit{Non-Convex Domain Case with Non-Linear Prices of both $ X_{1} $ and $ X_{2} $}. These are treated as $ \mathcal{D} =\{D_{1},D_{2}, D_{3}\}$ or $ \mathcal{D}\triangleq\{D_{j} | j=1,2,3\}  $. This is not a problem of ranking these domains to maximize revenue; rather, it is a choice of a revenue-maximizing domain subject to conditional purchase probability and demand uncertainty. These three domains are made of two commodities and lie in the $ j^{th} $ domain $ D_{j} $.  I assume that the probability of domain $ j $ being satisfactory is independent of everything else and denote it by $ \lambda_{j} $, i.e., conditional on viewing the two products on domain $ D_{j} $, the probability that she (customer) would purchase on this domain $ D_{j} $. 
Hence, the first task done here is to derive the demand function of each commodity $ X_{1}\& X_{2} $ under each domain $ D_{j} $ using the above four factors affecting the demand functions and after that using this demand function, and the purchase probability revenue maximization conditions and strategy have been discussed.\\
\indent The derivation of the perceived demand function and the customized offer function have been derived in three ways. The assumption is that the consumer buys two commodities $ X_{1} \& X_{2} $ to maintain the simplicity. The three cases are (i) Convex Domain Case with Linear prices of both $ X_{1} $ and $ X_{2} $, (ii) Convex Case with Linear $ X_{1} $ price, and Non-Linear  $ X_{2} $ Price and (iii) Non-Convex Domain Case with Non-Linear prices of both $ X_{1} $ and $ X_{2} $.
\subsection{Membership Function and Two-Way Consistency Condition}
\begin{figure}[h]
	\centering\includegraphics[scale=0.30]{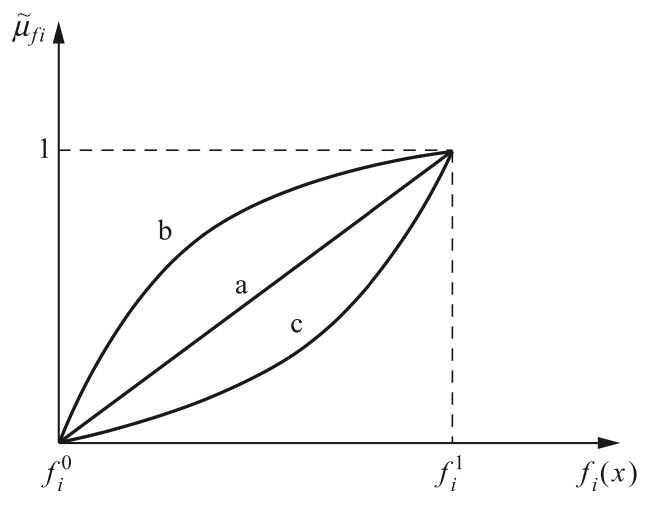}\caption{Fuzzy membership Functions of Different Forms}\label{figure2}
\end{figure}

The present model is based on a suitable choice of Membership Functions in understanding consumer behavior. The consumer motive is captured here by the membership function. Among numerous forms of fuzzy membership functions, three are commonly used: linear, parabolic, or reversed parabolic, as shown in Figure \ref{figure2}, and these are explained in \cite{fan2019mathematical}. In Figure \ref{figure2}, curve $ a  $ is the linear fuzzy membership function, representing the reasonable decision-maker. Curve $ b $ is the reversed parabolic function for the conservative decision-maker, and curve $ c $ is the explanatory function for the adventurous decision-maker. Regarding neoclassical economic theory, linear fuzzy membership represents risk-neutral behavior, reversed parabolic function describes risk-averse behavior, and parabolic function represents risk-taking behavior.
When solving standard economic problems, the linear fuzzy membership function shaped like curve $ a $ is adopted, mainly to simplify the mathematical description. Although the forms of fuzzy membership functions are different, the basic ideas are consistent. The linear membership function is generally expressed as:\\
\begin{equation*}\label{1}
	\phi_{f_{i}}(x)=\dfrac{f_{i}(x)-f_{i}^{0}}{f_{i}^{1}-f_{i}^{0}}
\end{equation*}
and \begin{equation}\label{2}
	\begin{aligned}
		& 	\phi_{f_{i}}(x)=0
		& \text{if}
		& & f_{i}(x)\leq f_{i}^{0}\\
		&	\phi_{f_{i}}(x)=1
		& \text{if}
		& & f_{i}(x)\geq f_{i}^{1}
	\end{aligned}
\end{equation}
In Eq. \ref{2}, $ f_{i}^{0}  $ is at an unacceptable level, whereas $ f_{i}^{1} $ is at a fully desirable level; the fuzzy membership function 
$\phi_{f_{i}}(x) $ reflects the membership degree of function $ f(x) $ between $ f_{i}^{0} $ and $ f_{i}^{1} $.
The membership degree can be considered an actual number from the interval $  [0, 1]$.
\paragraph{Two-way consistency of the choice problem.} The present study is based on the condition of rationality and is due to \cite{das2022understanding}.
\textit{There exists a two-way consistency of the choice problem. This means the choice would be the same if the decision-maker moved from a large set to a small one and from a small one to a large one. This happens provided the decision-maker can interpret the information correctly for each set}.\\
\indent The above definition indirectly says that the decision-maker is deciding with respect to a reference. It can be achieved in both ways, from a large or a small set, if the reference point is in both sets.\\
\indent This rationality condition is supported by the situation given by Arrow\cite{arrow1959rational}. It says that if some elements are chosen out of the set, say $ S_{2} $ and then the range of alternatives is narrowed to $ S_{1} $ but still contain some previously chosen elements, no previously unchosen piece becomes chosen, and no one selected part becomes unchosen. 
Let the set of alternatives are in $ \textbf{S}=\{x_{1},...,x_{n}\} $. Therefore, let any choice problem $ 	C: S_{i}\rightarrow \textbf{S} $ where, $ S_{i}\subseteq 2^{n} $. $ 2^{n} $ is the set of all subsets. The preference relation on $ \textbf{S} $ is reflexive, transitive, and asymmetric, and the relation is $ R=\textit{"better than or equal to"}i.e.\succcurlyeq $.
The choice problem is any subset $ S_{i} $, and the choice function is $ C_{\succcurlyeq}(S_{i})\rightarrow \textbf{S} $.  Let there are two choice problems; $ S_{1}\& S_{2}\subseteq\textbf{S} $.
This two-way consistency can be found in the degree of preference being constant in two sets. From Proposition 1 in \cite{das2022understanding},
\textit{If the agent maintains a constant preference in two-choice sets, the choice problem will follow a two-way consistency.}\\
let $ S_{1} $ and $ S_{2}$ are two subsets from $ S $. Then, the following two Equations are true.
\begin{equation*}\label{key}
	\forall S_{1}\subseteq S_{2}\subseteq \textbf{S}\text{if}C_{\succcurlyeq}(S_{2})\in S_{1}\text{then} C_{\succcurlyeq}(S_{1})=C_{\succcurlyeq}(S_{2})
\end{equation*}  
\begin{equation*}\label{key}
	\forall S_{1}\subseteq S_{2}\subseteq \textbf{S}\text{if}C_{\succcurlyeq}(S_{1})\in S_{2}\text{then} C_{\succcurlyeq}(S_{2})=C_{\succcurlyeq}(S_{1})
\end{equation*}
Moreover, information in the pattern $ S_{1} $ is equal to $ S_{2} $. This means,
\begin{equation*}\label{key}
	[C_{\succcurlyeq}(S_{1}|H)]= [C_{\succcurlyeq}(S_{2}|H)]=[x_{i}|H]
\end{equation*}
This means the agent selects the object $ x_{i}\in S $ when there were only $n$ objects and appear sequentially in ascending order in pattern $ S_{1} $and also selects the same thing $ x_{i}\in S $ when $ n $ and appear sequentially in descending order in pattern $ S_{2} $. In other words, the agent could correctly measure the information index/degree of preference for the object $ x_{i}\in S $ as it would happen in pattern $ S_{2} $.
If these two choices are the same, the agent would be two-way consistent. Here, $ H $ is the degree of preference, and it is the same in both subsets $ S_{1}\& S_{2}\subseteq S $, for example.
\subsubsection{Framework}
The model is based on some important variables. These are, $  X_{1},X_{2}\in\mathbb{R}_{+}$. Lowercase (viz. $ x_{1}\& x_{2} $) indicates the quantity in real numbers. In the present section, the demand for $ X_{1} $ and $ X_{2} $ of a consumer and in aggregate (industry) has been derived theoretically without knowing the individual consumers' utility functions. Therefore, the domain of the decision making is $ Z(=X_{1}\times X_{2})\subseteq \mathbb{R}_{+}^{2}$. The price per unit of $ X_{1} $ is $ p_{1} $ and the price per unit of $ X_{2} $ is $ p_{2} $. The total expenditure for the two variables is $ m $. A degree of preference towards the variable $ x_{1} $ is denoted by a membership function $ \mu\in\mathbb{R}_{+} $. Such that $ \mu:X_{1}\rightarrow [0,1] $. A degree of preference towards the variable $ x_{2} $ is denoted by a membership function $ \phi\in\mathbb{R}_{+} $. Such that $ \phi:X_{2}\rightarrow [0,1] $. $ x_{1min} $ is the minimum of $ X_{1} $ required to maintain the minimum quality of $ X_{2} $. $ x_{2min} $ is the minimum $ X_{2} $ required to maintain the quality of the goods and services in the market. $ x_{1max} \& x_{2max}$ are the variables that indicate the maximum consumption can be achieved under $ C $ for $ X_{1} $ and $ X_{2} $, respectively. The framework and the analysis here are based on the robust microeconomics framework given by Anwar Shaikh \cite{shaikh2016capitalism}. First, the demand function of $ X_{1} $ and $ X_{2} $ for an individual consumer will be derived. After that, the aggregate demand function will be derived. 
\subsubsection{Individual Demand Functions of $ X_{1} $ and $ X_{2} $}
For any price vector $ p $,the input vector $ z $ and the expenditure $ m $, the set of attainable alternatives is: $ B(p,m)=\{z\in Z\subseteq \mathbb{R}_{+}^{2}| pz\leqslant m\} $. A subset of $ Z $ having the form $ B(p,m) $ is called the budget set. One can obtain another available set by modifying the price $ p $ and the expenditure $ m $. Hence, $\mathcal{B}$ is treated as the family of the budget sets. Therefore, the choice/demand function is $ f:\mathcal{B}\rightarrow Z $.\\
\indent The derivation of the demand functions is divided into two parts, in an (i)Convex domain and an (ii) Non-Convex domain. The price will be linear if the market has high competition. This case would be treated as a convex case. Moreover, if the $ X_{2} $ product/service producers compete imperfectly and set the price non-linearly, this case would be treated as a non-convex case. The derivation of the individual demand function is based on the two-way consistency condition, as stated above. That considers that the consumer would maintain the previous proportion or the degree of preference towards a commodity. The fundamental intuition behind this model is that customers will continue to purchase the same product unless a trigger event forces them to consider other products. This may be termed as the motive of the consumer in our empirical measurement.  The detailed derivation of the proposed method can be found in \cite{das2022understanding,das2022t}. Hence, the demand functions can be derived in simple language by solving the following Equations individually for the two commodities $ X_{1} \& X_{2} $ respectively.
\begin{equation*}\label{key}
	\mu=\dfrac{(x_{1}-x_{1min})}{(x_{1max}-x_{1min})}
\end{equation*}
so that $ 0\leqslant \mu \leqslant 1 $.
Here, I assume for the time being that both $ x_{1min} $ and $ \mu $ are independent of prices. Then, for each $ \mu $, we can derive the corresponding per capita consumption demand for $ x_{1} $. The logic is also the same for $ X_{2} $. 
\begin{equation*}\label{3}
	\phi=\dfrac{(x_{2}-x_{2min})}{(x_{2max}-x_{2min})}
\end{equation*}
The analysis in the next three subsections assumes that  $ m $=\rupee 1000.
\subsubsection{Convex Domain Case with Linear prices of $ X_{1} \& X_{2} $ Commodities}
\begin{figure}[h]
	\centering\includegraphics[scale=0.30]{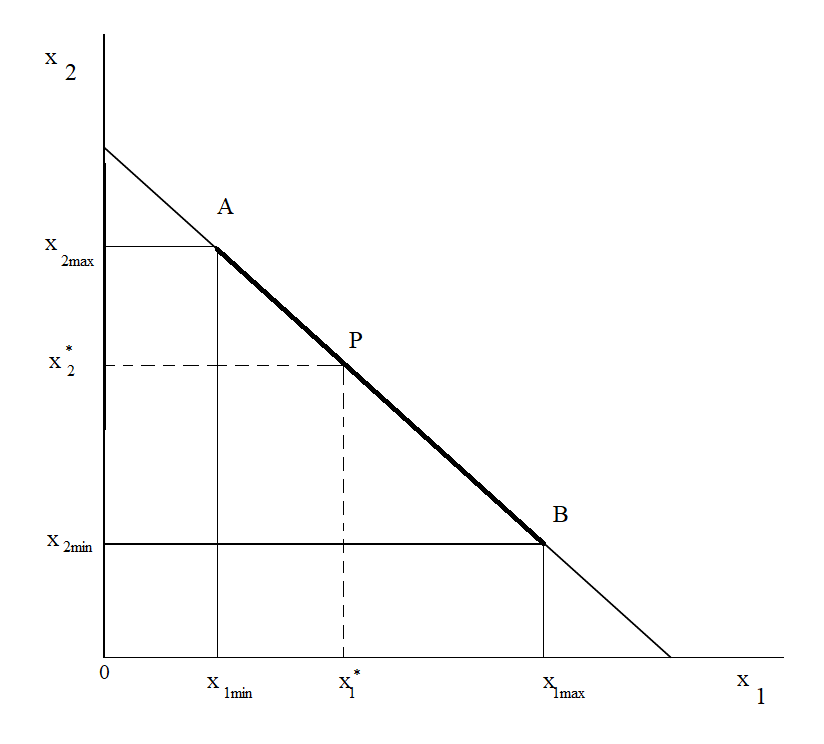}\caption{Convex Domain Analysis}\label{figure44}
\end{figure}
\begin{figure}[H]
	\centering\includegraphics[scale=0.15]{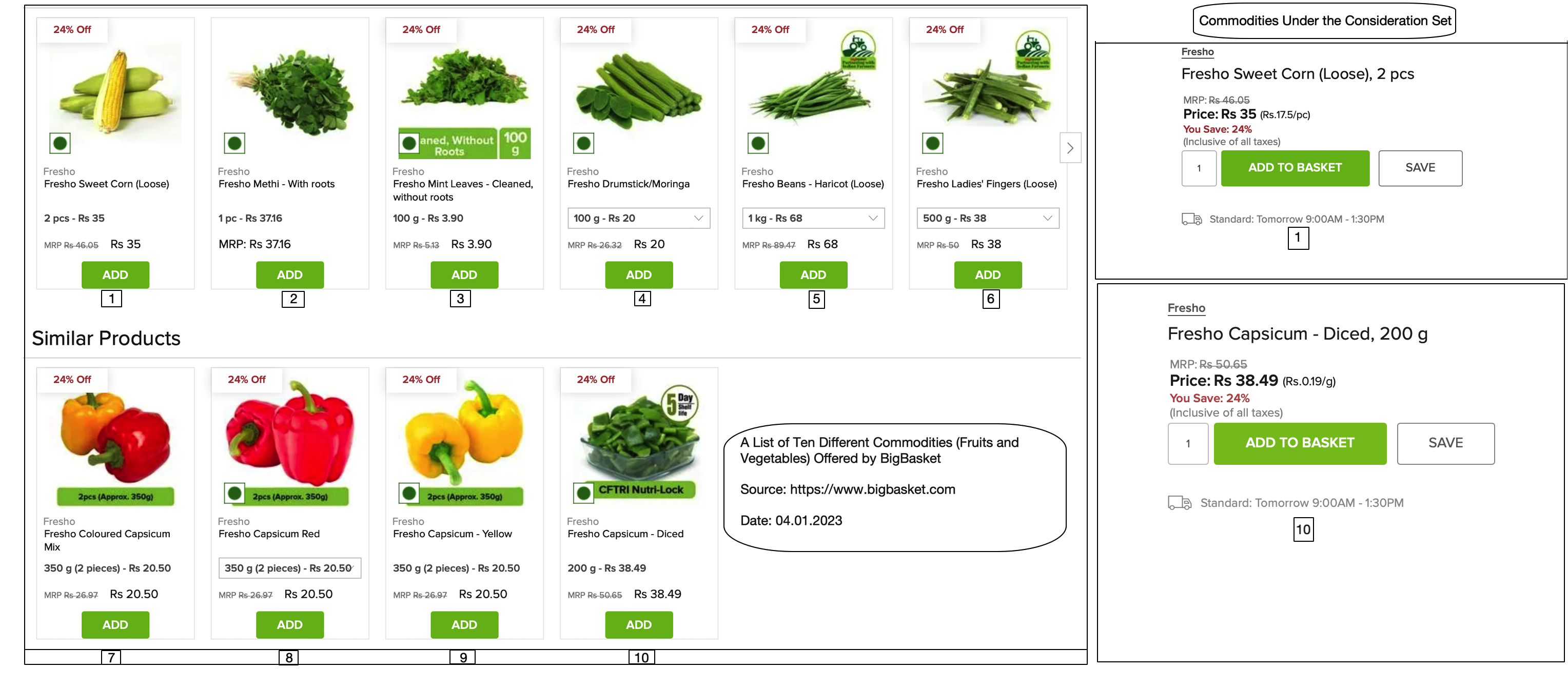}\caption{Set of Ten Different Types of Products offered by BigBasket along with the Consideration Set-One of the Consumer}\label{figure31}
\end{figure}
\begin{figure}[H]
	\centering\includegraphics[scale=0.20]{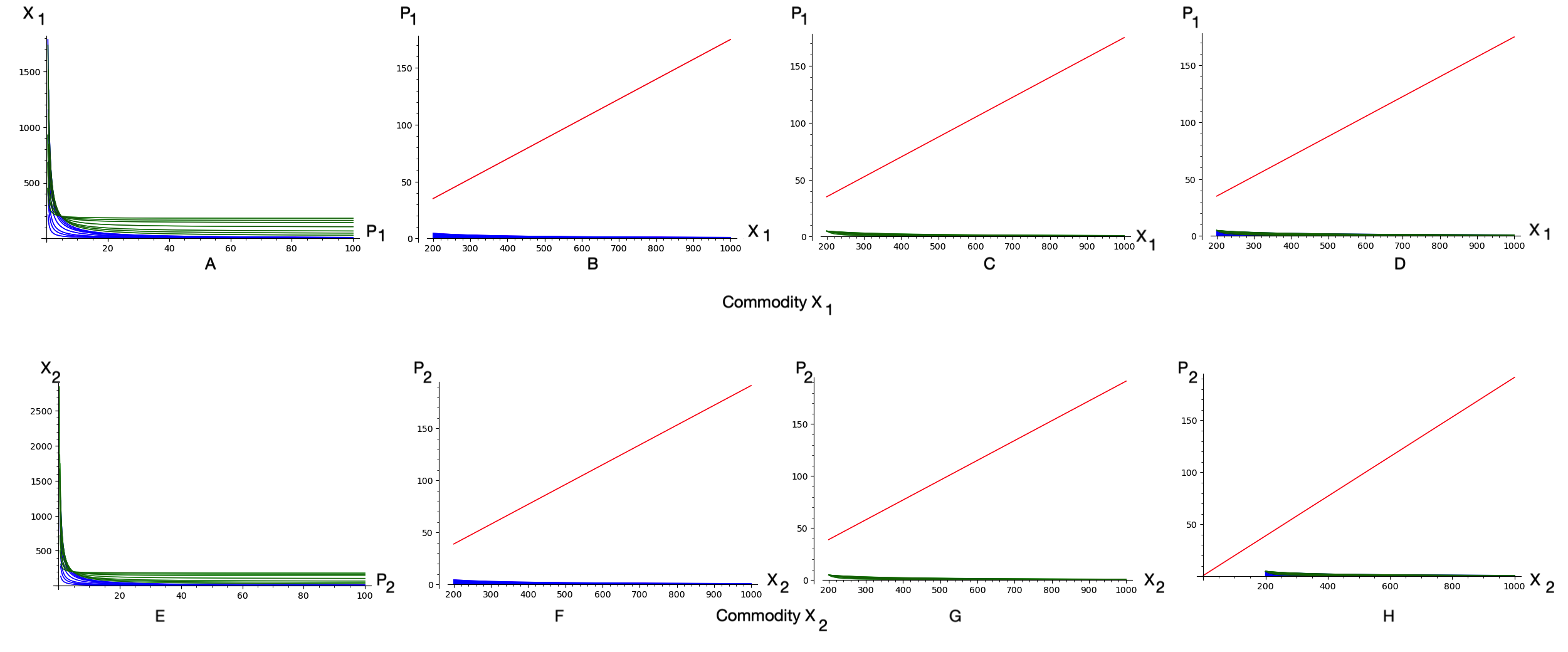}\caption{Numerical Analysis of Convex Domain Case with Linear $ X_{1} $ price and Linear  $ X_{2} $ Price}\label{figure39}
\end{figure}

Figure \ref{figure44} depicts the Convex Domain Analysis case.
Let Figure \ref{figure31} be any offer by an online retail platform of fruits and vegetables. This is the situation where in the consideration set from Figure \ref{figure31}, commodity [1] and [10] are treated $ x_{1} \& x_{2}$ respectively.
\begin{equation}\label{4}
	x_{1}=(1-\mu)x_{1min}+\mu(\dfrac{m}{p_{1}})-\mu(\dfrac{p_{2}}{p_{1}})x_{2min}	
\end{equation}
\begin{equation}\label{5}
	x_{2}=(1-\phi)x_{2min}+\phi(\dfrac{m}{p_{2}})-\phi(\dfrac{p_{1}}{p_{2}})x_{1min}	
\end{equation}
The two Equations \ref{4} \& \ref{5} are the demand equations of $ X_{1} $ and $ X_{2} $ respectively\footnote{The derivations are in the appendix}. It is apparent from Equations \ref{4} \& \ref{5} that for each good, the quantity demanded responds negatively to a rise in its price at any given income (Figure \ref{figure39} in A and E). Yet, it requires no specific model of consumer behavior. \\
\indent From Figure \ref{figure31}, the consideration set consists of two commodities, and we consider the following values for the parameters: $ p_{1}=35;p_{2}=38.49;x_{1min}= 2pcs;x_{2min}=200g,m=1000$. The parameters $ \mu,\phi $ are unknown; we have taken three different values to measure the mismatch. These are $ \mu,\phi=\{0.10,0.20,0.30,0.50,0.70,0.80,0.9\} $ as the least preferred to highly preferred ranges. The commodity purchase ranges are $ x_{1}=\{2,4,6,8,...\} \& x_{2}=\{200,400,600,800,...\}$ respectively and are in $ \mathbb{Z}_{+} $. But to derive the perceived demand function, we need to assume that $ x_{1},x_{2}\in \mathbb{R}_{+}$. This means the domain is linear i.e. $ 0.175x_{1}+0.19x_{2}=1000 $ assuming $ x_{1}, x_{2}\in \mathbb{R}_{+} $. Here, $ p_{1}=0.175/\text{per g} \& p_{2}=0.19/g $. Hence, assuming the minimum order quantity as $ 1 pcs$ for $ x_{1} $ and $ 1g $ for $ x_{2} $, the consumer's motive as given by $ \mu \& \phi $, linear or convex domain the two demand functions can be derived as below. Here, we also assume that 2pcs =200g.
\begin{equation}\label{6}
	x_{1}=(1-\mu)+\mu(\dfrac{1000}{p_{1}})-\mu(\dfrac{0.19}{p_{1}})
\end{equation}
\begin{equation}\label{7}
	x_{2}=(1-\phi)+\phi(\dfrac{1000}{p_{2}})-\phi(\dfrac{0.175}{p_{2}})
\end{equation}
And if considering the minimum order of $ x_{1} \& x_{2}$, the demand functions are given below.
\begin{equation}\label{8}
	x_{1}=(1-\mu)200+\mu(\dfrac{1000}{p_{1}})-\mu(\dfrac{0.19}{p_{1}})200
\end{equation}
\begin{equation}\label{9}
	x_{2}=(1-\phi)200+\phi(\dfrac{1000}{p_{2}})-\phi(\dfrac{0.175}{p_{2}})200
\end{equation}
On the other hand, the offered pair set or the supply schedule is given by $\{(35,200g),(70,400g),(105,600g),\\...\}$ for $ x_{1} $ and $ \{(38.9,200g),(76.98,400g),(115.47,600g),...\} $ for $ x_{2} $. Using these data pairs, the supply equation can be derived as $ p_{1s}=0.175x_{1} \& p_{2s}=0.1904x_{2}+0.8199999999999932$ for $ x_{1} \& x_{2} $ respectively.\\
\textbf{Remarks 1} For all values of $ \mu \& \phi $ and $ p_{1} \& p_{2}  $ the demands of $ x_{1} \& x_{2} $ are higher in presence of minimum order constraints in Equations \ref{8} \& \ref{9} than in Equations \ref{6} \& \ref{7} respectively. As a result, the revenue will be higher in the presence of minimum order constraints. This minimum order pair is a part of the domain. Hence, the equilibrium price will be more to maximize the revenue as proposed in Equation \ref{4} \& \ref{5}.\\
\textbf{Numerical Results}
In the following simulated result Figure \ref{figure39}, we have shown that the equilibrium price is higher in the revenue functions of commodity $ x_{1}\& x_{2} $ in the presence of the minimum purchase constraints. On the other panel, we have shown that demand and supply are equal at that equilibrium price. There are two Panels. In Panel 1, the analysis for commodity $ x_{1} $ has been done, and in Panel 2, for commodity $ x_{2} $. In both cases, the results are the same. Diagrams A and E show two curves for $ x_{1}\& x_{2} $ respectively. Blue curves are based on Equations \ref{6} \& \ref{7}. While green curves are based on \ref{8} \& \ref{9}. The curves are drawn assuming $ \mu \& \phi =\{0.1,0.2,0.3,0.4,0.5,0.6,0.7,0.8,0.9\} $. The demands curve is a high degree of $ \mu \& \phi $ higher. Moreover, demand increases and shifts above in the presence of minimum order constraints.  This can be seen that the green curves are above the blue curve. The other Diagrams are drawn by changing the quadrant. Combining B and C, D was drawn, and F, G, and H were combined. The red line is the platform's supply line. The equilibrium quantity will be determined by considering the aggregate demand not depicted here. The platform is the price taker. So, the Diagram shows that the equilibrium quantity will be meager because the demand curves show a very high degree of elasticity. This is because the offered commodities are necessary goods. Here, in the aggregate demand functions, the parameters $ \mu \& \phi $ are nothing but the maximum of the individual firm's preferences for $ X_{1} $ and $ X_{2} $.\\ 
\textbf{Aggregate Demand Functions.} Here, in the aggregate demand function\footnote{The derivations are in the appendix}, the parameters $ \mu \& \phi $ are nothing but the maximum of the individual customer's preferences for $ X_{1} $ and $ X_{2} $. 
\begin{equation}\label{10}
	\sum_{i=1}^{N}x_{1_{i}}=(1-\mu_{k})	\sum_{i=1}^{N}x_{1min_{i}}+	(\dfrac{\mu_{k}}{p_{1}})\sum_{i=1}^{N}m_{i}-\mu_{k}(\dfrac{p_{2}}{p_{1}})\sum_{i=1}^{N}x_{2min_{i}}	
\end{equation}
\begin{equation}\label{11}
	\sum_{i=1}^{N}x_{2_{i}}=(1-\phi_{k})\sum_{i=1}^{N}x_{2min_{i}}+(\dfrac{\phi_{k}}{p_{2}})\sum_{i=1}^{N}m_{i}-\phi_{k}(\dfrac{p_{1}}{p_{2}})\sum_{i=1}^{N}x_{1min_{i}}	
\end{equation}
The industry faces the same prices, viz. $ p_{1}\& p_{2} $, but different budget i.e. $ C_{i} $.Moreover, $ x_{1min_{i}},x_{2min_{i}} $ are also different for the different customers.The aggregate $ \mu $ is $ \mu_{k}=\{\mu_{1}\cup \mu_{2}\cup...\cup \mu_{N}\} =\max_{i=1}^{N}\mu_{i}$ and the aggregate $ \phi $ is $ \phi_{k}=\{\phi_{1}\cup \phi_{2}\cup...\cup \phi_{N}\} =\max_{i=1}^{N}\phi_{i}$ .The same thing is true for $ (1-\mu_{k})=\max_{i=1}^{N}(1-\mu)_{i}$ and $ (1-\phi_{k})=\max_{i=1}^{N}(1-\phi)_{i}$. Because, $ \mu+(1-\mu)=1 \& \phi+(1-\phi)=1 $. Moreover, $ \mu,\phi\in[0,1]\subseteq\mathbb{R}_{+} $.
The platform will achieve maximum revenue because of the high volume of aggregate minimum orders and the high degree of aggregate preferences $ \mu_{k} \& \phi_{k} $.
\subsubsection{Convex Case with Linear $ X_{1} $ price and Non-Linear  $ X_{2} $ Price}
\begin{figure}[h]
	\centering\includegraphics[scale=0.50]{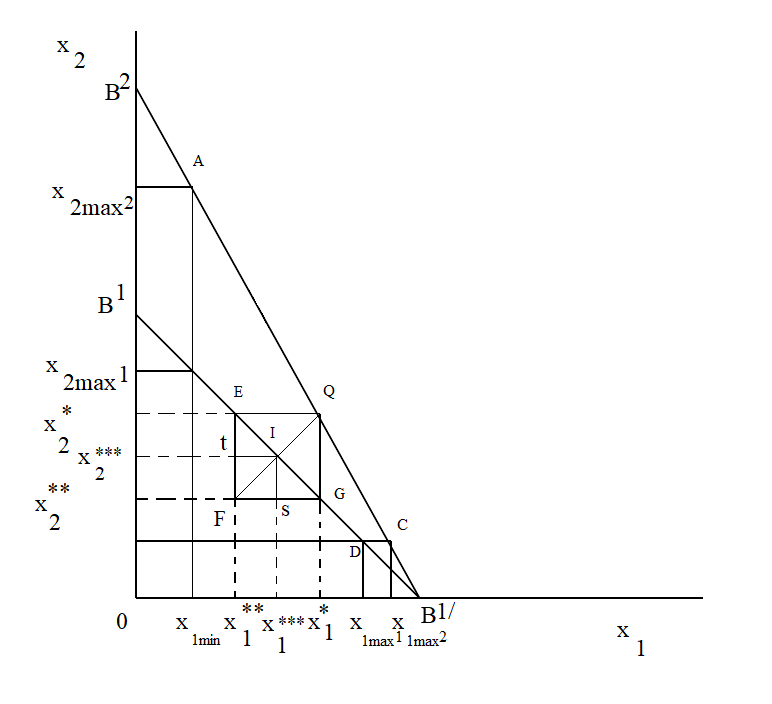}\caption{Convex Case with linear $ X_{1} $ price and Non-Linear $ X_{2} $ Price:(i)}\label{figure46}
\end{figure}
\begin{figure}[h]
	\centering\includegraphics[scale=0.50]{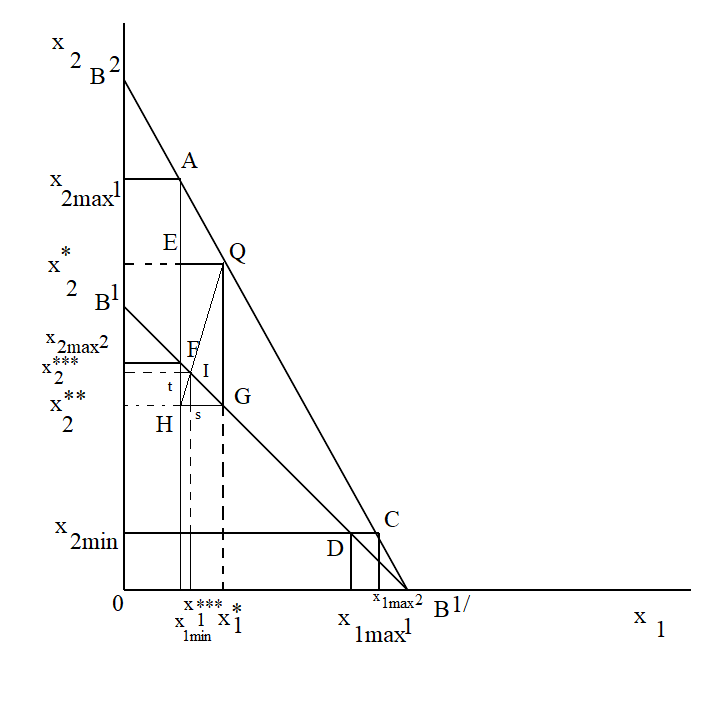}\caption{Convex Case with linear $ X_{1} $ price and Non-Linear $ X_{2} $ Price:(ii)}\label{figure47}
\end{figure}
Two possibilities are depicted in \ref{figure46} \& \ref{figure47}. Derivations are given in the appendix.
\begin{figure}[H]
	\centering\includegraphics[scale=0.15]{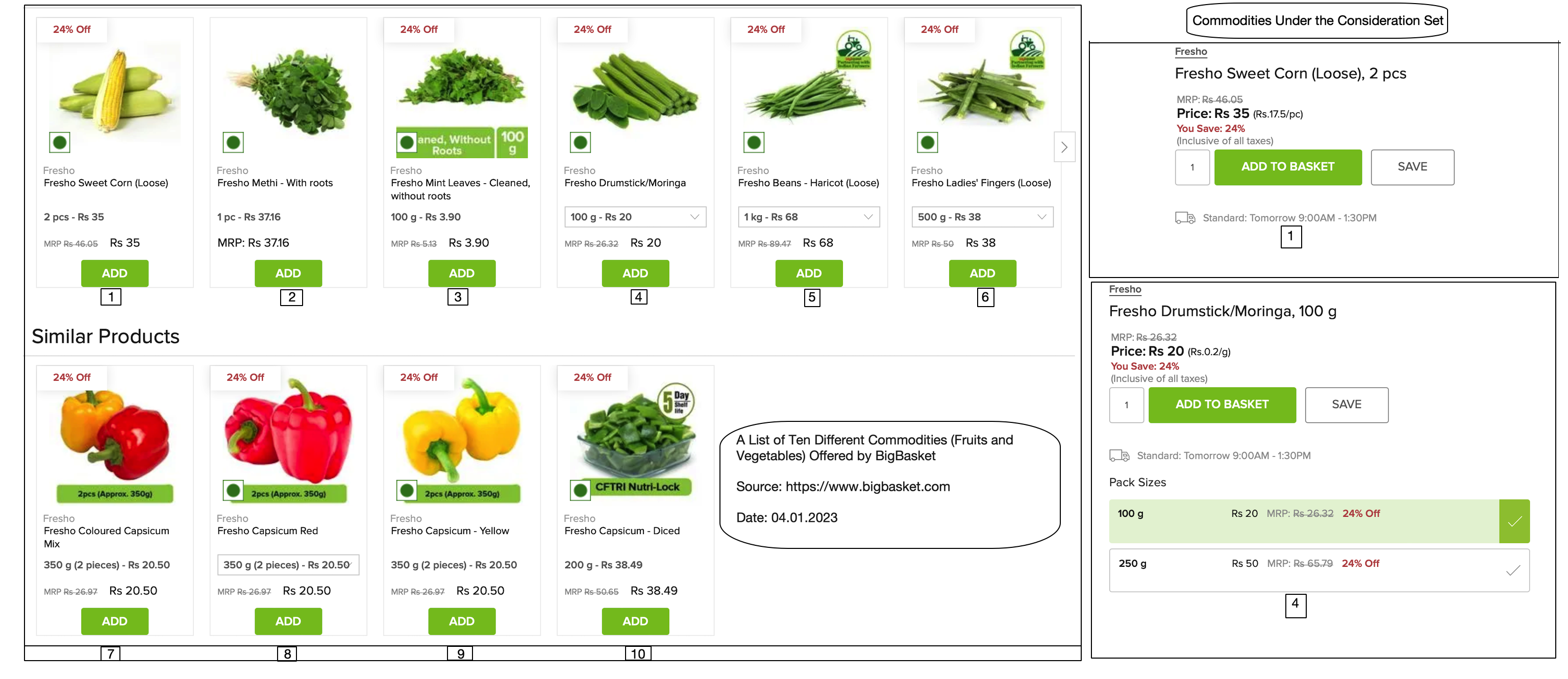}\caption{Set of Ten Different Types of Products offered by BigBasket along with the Consideration Set-Two of the Consumer}\label{figure32}
\end{figure}
Let another offer by any platform is represented in Figure \ref{figure32}. This is the situation where in the consideration set from Figure \ref{figure32}, [1] and [4] are treated $ x_{1} \& x_{2}$ respectively. The offered price for $ x_{2} $ follows a step function and is nonlinear (see Figure \ref{figure34}). As a result, the domain is non-convex. For a non-convex domain, the solution will be either at the corner or in the interior of the feasible region based on the constraints. As a result, the revenue will be more than in the previous case. In this situation, the price of commodity $ x_{1} $ is linear, but the price of commodity $ x_{2} $ is non-linear so that the solution will be in the interior\cite{das2022understanding,  das2022t}. Numerical parameters are the same for $ x_{1} $ but different for $ x_{2} $. $ p_{2}^{1}=0.2 $ for $ x_{2min}=100g $ but after $ 100g $ the next quantity is more than $ 200 $, i.e., $ 250 $. As a result, the gap between the perceived and the offered demand curve will be more than in the previous case. Now, if the $ p_{2}^{1}=0.1 $ for $ x_{2min}=200g $, then the domain would be non-convex, and the following demand functions would persist. Let the vector $ z=(x_{1},x_{2}) $ for two variables and,$ z\in Z\subseteq \mathbb{R}_{+}^{2} $.There are two budget sets that the firm is facing.These are, $ B^{1}=\{z_{i}\in\mathbb{R}_{+}^{2}:\sum_{i=1}^{2}p_{i}^{1}x_{i}\leq m^{1}\}\&  B^{2}=\{z_{i}\in\mathbb{R}_{+}^{2}:\sum_{i=1}^{2}p_{i}^{2}x_{i}\leq m^{2};\text{if},x_{2}>\bar{x}_{2} \}$. Moreover,$ m^{2}>m^{1},p_{1}^{1}\neq p_{1}^{2} \& p_{2}^{1}<p_{2}^{2}$, because if you buy more of $ x_{2}$, you have to pay more and as a result, the relative price of $ x_{1} $ will also change. Here, $ \bar{x}_{i} \text{for},i=1,2$ are any constant real numbers of $ x_{1}\& x_{2} $.The different prices for these constraints create the price non-linearly\footnote{The derivations are in the appendix}.
\begin{equation}\label{12}
	x_{1}=\dfrac{\mu_{i}}{p_{1}^{1}}[(m^{1}-p_{2}^{1}x_{2min})-p_{1}^{1}x_{1min}]+x_{1min}
\end{equation} 
\begin{equation}\label{13}
	x_{2}=\dfrac{\phi_{i}}{p_{2}^{1}}[(m^{1}-p_{1}^{1}x_{1min})-p_{2}^{1}x_{2min}]+x_{2min}
\end{equation} 
\textbf{Remarks 2}
The price of commodity $ x_{2} $ is higher in the higher quantity buy, so the solution will be at the lowest quantity. This is why $ p_{2}^{2}, m^{2}, p_{1}^{2} $ are not there in the above two equations.  This can be checked from Figure \ref{figure32}  that the consumer would buy 100g at  \rupee $ 0.20 $. It is also clear from Equations \ref{12} \& \ref{13}\footnote{The derivations are in the appendix} that whatever the price, there is a minimum demand for each commodity.\\
\textbf{Aggregate Demand Functions.}Here, is the aggregate demand functions for $ X_{1} $ and $ X_{2} $.\footnote{The derivations are in the appendix} 
\begin{equation*}\label{14}
	\sum_{i=1}^{N}x_{1_{i}}=\mu_{k}[p_{1}^{1}\sum_{i=1}^{N}m_{i}^{1}-\dfrac{p_{2}^{2}p_{1}^{1}\sum_{i=1}^{N}m_{i}^{1}}{m^{2}}x_{2min_{k}}]+\mu_{k-1}[\dfrac{\sum_{i=1}^{N}m_{i}^{1}}{p_{1}^{1}}-\dfrac{p_{2}^{1}}{p_{1}^{1}}\sum_{i=1}^{N-1}x_{2min_{i\neq k}}]+(1-\mu_{k})\sum_{i=1}^{N}x_{1min_{i}}
\end{equation*}
and 
\begin{equation*}\label{15}
	\sum_{i=1}^{N}x_{2_{i}}=\phi_{k}[\dfrac{\sum_{i=1}^{N}m^{2}_{i}}{p_{2}^{2}}-\dfrac{\sum_{i=1}^{N}m_{i}^{2}}{p_{1}^{1}p_{2}^{2}\sum_{i=1}^{N}m_{i}^{1}}x_{1min_{k}}]+\phi_{k-1}[\dfrac{\sum_{i=1}^{N}m_{i}^{1}}{p_{2}^{1}}-\dfrac{p_{1}^{1}}{p_{2}^{1}}\sum_{i=1}^{N-1}x_{1min_{i\neq k}}]+(1-\phi_{k})\sum_{i=1}^{N}x_{2min_{i}}
\end{equation*}
Here, $ \sum_{i=1}^{N-1}\mu_{i\neq k}=\cup_{i=1\neq k}^{N-1}(\mu_{i\neq k})=\max[\mu_{1,}...,\mu_{N-1}]=\mu_{k-1} $ \\and $ \sum_{i=1}^{N-1}\phi_{i\neq k}=\cup_{i=1\neq k}^{N-1}(\phi_{i\neq k})=\max[\phi_{1},...,\phi_{N-1}]=\phi_{k-1} $ for an order $ \mu_{1}<...<\mu_{k-1}<\mu_{k} $ \\ and $ \phi_{1}<...<\phi_{k-1}<\phi_{k},[\mu_{k}+\sum_{i=1}^{N-1}\mu_{i\neq k}]=\max[[\mu_{k}]+\max_{i=1}^{N-1}\mu_{i\neq k}] =\mu_{k}$ and $ [\phi_{k}+\sum_{i=1}^{N-1}\phi_{i\neq k}]=\max[[\phi_{k}]+\max_{i=1}^{N-1}\phi_{i\neq k}] =\phi_{k}$.Moreover,$ x_{1min_k} \& x_{2min_k} $ are the minimum fixed demand for the firm $ k $. This case has proposed Proposition 1.
\begin{proposition}
	If the price of $ X_{2} $ in the market is in the non-linearly set whereas $ X_{1} $ is linearly priced, then it suggests that the suppliers of $ X_{2} $ would enjoy some degree of monopoly power. On the other hand, the $ X_{1} $ would enjoy a linear price. The firms who can use $ X_{2} $ at a higher cost would use it, and the rest would prefer $ X_{1} $. Hence, there would be multiple equilibria in the market. As a result, $ X_{1} $ and $ X_{2} $would act as complementary to each other.
\end{proposition}
\begin{proof}
	The proof is in the appendix.
\end{proof}
\subsubsection{Non-Convex Domain Case with Non-Linear $ X_{1} $ price and Non-Linear  $ X_{2} $ Price}
	\begin{figure}[h]
	\centering\includegraphics[scale=0.30]{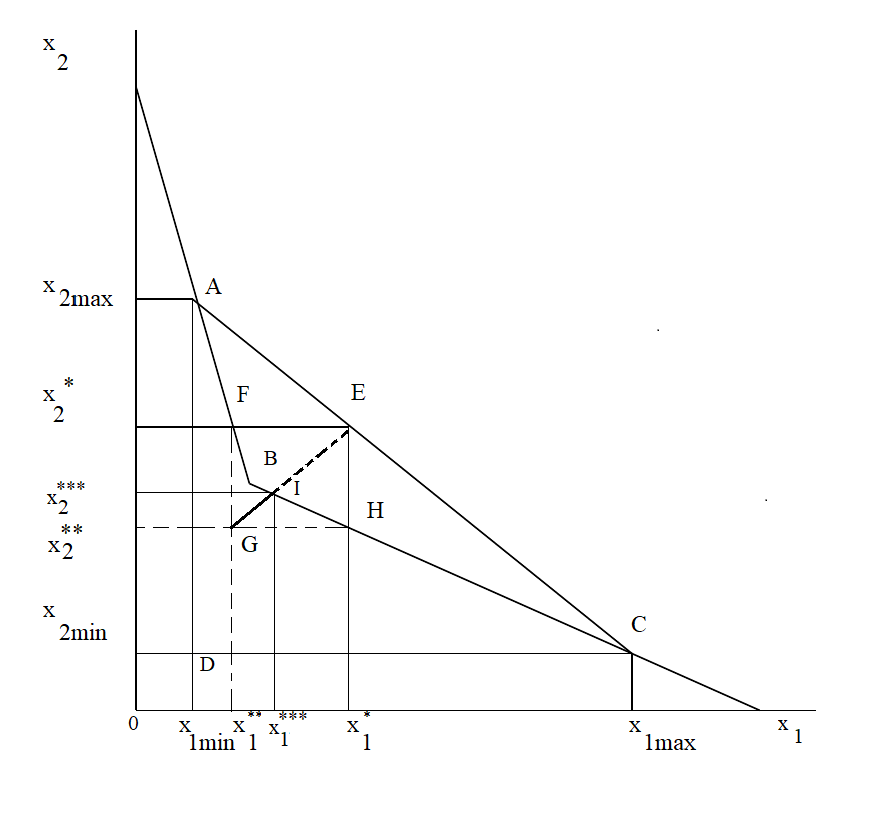}\caption{Non-Convex Domain Analysis}\label{figure45}
\end{figure}
Non-Convex Domain Case has been depicted in \ref{figure45}. 
\begin{figure}[H]
	\centering\includegraphics[scale=0.15]{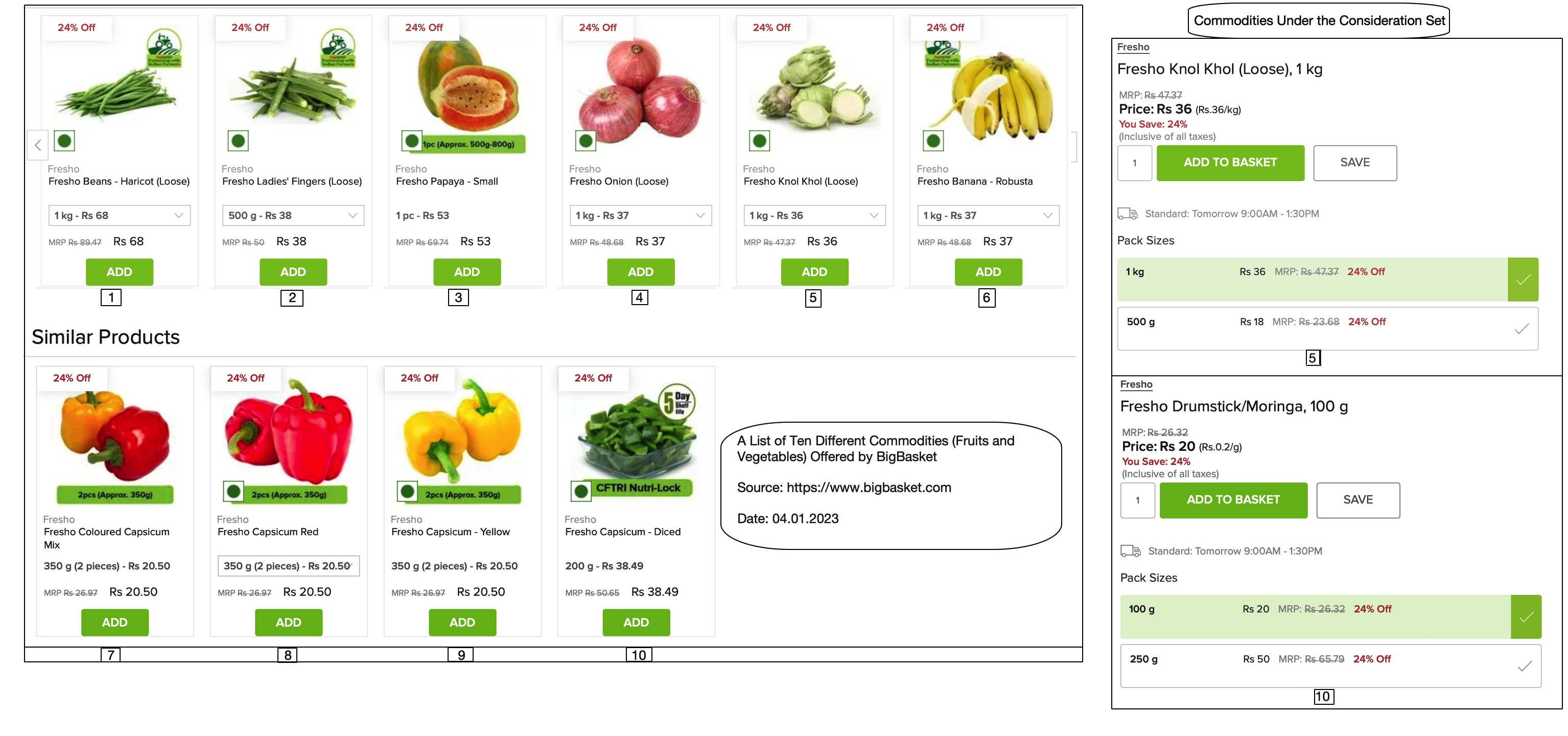}\caption{Set of Ten Different Types of Products offered by BigBasket along with the Consideration Set-Three of the Consumer}\label{figure33}
\end{figure}
\begin{figure}[H]
	\centering\includegraphics[scale=0.20]{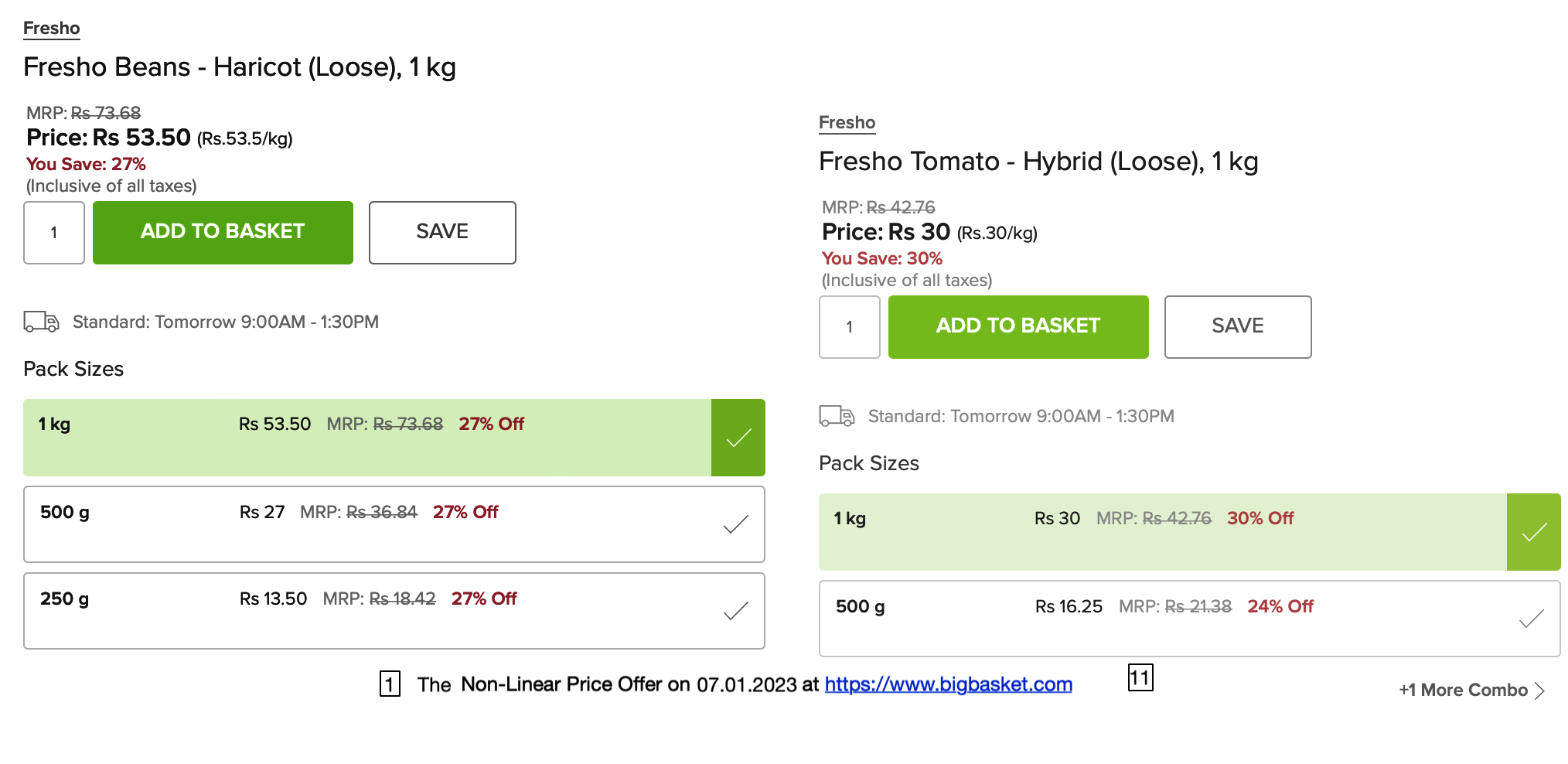}\caption{The Non-Linear Price Offer}\label{figure34}
\end{figure}
This is the situation where in the consideration set from Figure \ref{figure34} commodity [1] \& [11] (from Figure \ref{figure34})and either [5] or [10] (from Figure \ref{figure33}) is treated as $ x_{1} \& x_{2}$ respectively. The offered price for $ x_{1} \& x_{2} $ follows a step function and is nonlinear. For example, in Figure \ref{figure34}, the price per unit is  \rupee 0.0540, and the minimum quantity bought is 250g. If the consumer buys at least 1kg or 1000g, the per unit price will reduce to \rupee 0.05350/g. Hence, the minimum quantity bought is 1000g or 1kg. This is a complete non-convex domain, and the demand functions for these two commodities can be derived below. Let the vector $ z=(x_{1},x_{2}) $ for two variables and,$ z\in Z\subseteq \mathbb{R}_{+}^{2} $.There are two budget sets that the firm is facing.These are, $ B^{1}=\{z_{i}\in\mathbb{R}_{+}^{2}:\sum_{i=1}^{2}p_{i}^{1}x_{i}\leq m;\text{if},  x_{1}\leq \bar{x}_{1} \}\&  B^{2}=\{z_{i}\in\mathbb{R}_{+}^{2}:\sum_{i=1}^{2}p_{i}^{2}x_{i}\leq m;\text{if},x_{1}>\bar{x}_{1}\& x_{2}<\bar{x}_{2} \}$.Here, $ \bar{x}_{i} \text{for},i=1,2$ are any constant real numbers of $ x_{1}\& x_{2} $. The firm faces two budget constraints at the same time, so the domain becomes non-convex because the price vectors $ (p_{1}^{1},p_{2}^{1})(p_{1}^{2},p_{2}^{2}) $ for the two budget sets are different and depends on the quantity purchase, i.e., $ \bar{x}_{i},i=1,2 $. Assuming $ (m-\bar{m}_{1}p_{1}^{2})=p_{1}, (m-\bar{m}_{2})p_{2}^{1}=p_{2}\& (m-\bar{m}_{1})(m-\bar{m}_{2})=M$. Assuming that the minimum expenditure in $ x_{1}\& x_{2} $ are constant i.e. $ p_{1}^{1}x_{1min}=\bar{m}_{1}, p_{2}^{2}x_{2min}=\bar{m}_{2} $. There are two possibilities viz. \textit{$ \bar{m}_{1}\neq\bar{m}_{2} $} \& 
\textit{$ \bar{m}_{1}=\bar{m}_{2}=\bar{m} $}.The different prices for these constraints create a non-linear price.  The derived demand functions\footnote{The derivations are in the appendix} are given below.
\begin{equation}\label{16}
	x_{1}=\dfrac{\mu}{p_{1}}[\dfrac{p_{1}}{p_{1}^{2}}(m-p_{2}^{2}x_{2min})-p_{1}x_{1min}]+x_{1mn}
\end{equation}
\begin{equation}\label{17}
	x_{2}=\dfrac{\phi}{p_{2}}[\dfrac{p_{2}}{p_{2}^{1}}(m-p_{1}^{1}x_{1min})-p_{2}x_{2min}]+x_{2min}
\end{equation}
Using Equations \ref{16} \& \ref{17} I derived as below.
\begin{equation}\label{18}
	x_{1}^{*}=\dfrac{m-p_{2}^{1}x_{2}^{}}{p_{1}^{1}}
\end{equation}
\begin{equation}\label{19}
	x_{2}^{*}=\dfrac{m-p_{1}^{2}x_{1}^{}}{p_{2}^{2}} 
\end{equation}
\textbf{Aggregate Demand Functions.}
The aggregate demand function\footnote{The derivations are in the appendix} can be found by summing for $ N $ number of consumers in the industry. First, we need to convert the demand functions \ref{16} \& \ref{17} into aggregate terms, as below. The aggregate demand function for $ X_{2} $ depends on the degree of preference of the large customer.
\begin{equation*}\label{20}
	\sum_{i=1}^{N}x_{1_{i}}^{*}=\dfrac{\mu_{k}}{p_{1}}[\dfrac{p_{1}}{p_{1}^{2}}(\sum_{i=1}^{N}m_{i}-p_{2}^{2}\sum_{i=1}^{N}x_{2min_{i}})-p_{1}\sum_{i=1}^{N}x_{1min_{i}}]+\sum_{i=1}^{N}x_{1mn_{i}}
\end{equation*} 
and 
\begin{equation*}\label{21}
	\sum_{i=1}^{N}x_{2_{i}}^{*}=\dfrac{\phi_{k}}{p_{2}}[\dfrac{p_{2}}{p_{2}^{1}}(\sum_{i=1}^{N}m_{i}-p_{1}^{1}\sum_{i=1}^{N}x_{1min_{i}})-p_{2}\sum_{i=1}^{N}x_{2min_{i}}]+\sum_{i=1}^{N}x_{2min_{i}}
\end{equation*}
Finally, using Equations (25) and (26), the final range of the demand equations can be found.These are; $ (\sum_{i=1}^{N}x_{1_{i}},\sum_{i=1}^{N}x_{2_{i}} )\in[(\sum_{i=1}^{N}x_{1_{i}}^{**},\sum_{i=1}^{N}x_{2_{i}}^{**})],(\sum_{i=1}^{N}x_{1_{i}}^{***},\sum_{i=1}^{N}x_{2_{i}}^{***})$.\\
\textbf{Remarks 3.}
The interpretation is very clear from the Equations \ref{16}, \ref{17}, \ref{18}, \ref{19}, i.e., there is a two-way complementary relationship that exists between the two commodities, viz. $ x_{1} \& x_{2}$. Say, for example, for commodity $ x_{1} $. In the first round, the demand depends on the minimum expenditure to buy $ x_{2} $. The higher the minimum expenditure incurred, the lower the demand for $ x_{1} $. Now, if all the parameters are fulfilled and decide $ x_{1} $ in \ref{16}, the next level or the final decision will be taken in Equation \ref{18}. Again, the demand of $ x_{2} $ matters based on Equation \ref{13}. Higher is the demand of $ x_{2} $ lower is the demand of $ x_{1} $. The same analysis is true for $ x_{2} $. Therefore, the higher the non-linearity of the step functions, the lower the demand, and the consumer would not prefer that domain to buy. Rather, they prefer to buy from a domain with a linear domain. The choice of domain and the conditions thereof have been explained in the following section as an application toward consumer behavior. Thus far, it is clear that the platform will be able to earn more revenue if the domain contains step functions of price-quantity relationship or non-linearity. Because of this nonlinearity, the choice set becomes very small for the consumer. As a result, the consumer needs to spend more on that domain.\\
\textbf{Remarks 4.}
From the above model, it is clear that the minimum quantity constraints create the decision space nonconvex. Hence, the proposed method also suggests that the nonconvex domain does not guarantee that the solution will be at the optimum point.  Clearly and accurately, the consumer's decision-making domain is a lattice. All the parameters, viz. prices, discounts, minimum order quantities, etc... are in integer lines. As a result, the solution considered is in real space, and the nonconvex domain provides a set of solutions in the interior. Now, it is necessary to know how many lattice points there are in possible solutions. The individual demand function derived in equations \ref{4}\& \ref{5} predicts the demand quantities in real numbers for different minimum purchase constraints and other parameters. Hence, the second question is to identify the lattice commodity bundle nearest to the real one concerning the demand function defined in actual variables. The above questions can be solved in a process as follows. The estimated/projected demand vector in real numbers can be determined from the individual demand function. After that, a set of available lattice points will be determined. In the end, the nearest lattice points need to be derived and match the same. 
\paragraph{Remarks 5}
The revenue is dependent on the choice of the products offered, respected prices, the minimum order constraints, the motive of the customer as represented by the degree of preference for a product in different slabs of each offer, etc... So far, I have considered that the customer motive viz, $ \mu, \phi $ are identical in both linear and non-linear offers. This means that it was not dependent on slabs. For the sake of a clear understanding of the price response function, hereafter, the motive of the customer is different in the nonlinear offer for each slab so that the expected revenue can be calculated for a fixed and random attention span. This means each customer may have a different preference for various slabs of a given offer. Let us try to understand the using Figure \ref{9}. To buy Fresho Beans, there are three slabs in the left panel: viz. {(1 kg, @ \rupee 53.50),(500g, @\rupee 27),(250g, @\rupee13.50)}. So far, it was assumed that the degree of preference, say $ \mu $, to derive the demand function of this commodity is constant in these three slabs. To generalize, $ \mu $ is different for these three slabs. 
\section{The Price Response Function and its Properties}
This section explains the properties of the generalized price response function, as given by the following equation for commodity $ x_{i} $:
\begin{equation}\label{22}
	x_{i}(p_{i}\mid\mu^{k},x_{i,min}^{k},m,p_{j}^{k},x_{j,min}	^{k};\forall j,k=1,...,N)=(1-\mu^{k})x_{i,min}^{k}+\mu^{k}(\dfrac{m}{p_{i}^{k}})-\mu^{k}(\dfrac{p_{j}^{k}}{p_{i}^{k}})x_{j,min}	^{k}
\end{equation}
In economic textbooks, the price-response function is analogous to the market demand function. There is, however, a significant distinction. The price-response function describes the demand for a single seller's goods as a function of the price given by that seller. This contrasts the concept of a market demand curve, which outlines how a whole market will react to price changes. This distinction is significant because various firms competing in the same market confront different price-response functions \cite{phillips2021pricing,pinchuk2006applying}. Eq.\ref{22} gives the general demand function of commodity $ x_{i} \forall i=1,...,N$  and for all possible slabs $ k=1,...,N $ and for other commodity $x_{j} \forall j=1,...,N $. If the platform offers in the non-convex domain, then for each slab, the price response function can also be calculated using Eq.\ref{22} instead of calculating the overall response for that domain. Because each slab offer is linear, nonlinearity arises when considering a set of linear offers for all slabs. For example, in Figure \ref{9}, to buy Fresho Beans, there are three slabs; this is considered the set of offers viz. $ D_{1} $= {(1 kg, @ \rupee 53.50),(500g, @\rupee 27),(250g, @\rupee13.50)}. The set $ D_{1} $ is a nonlinear offer, but individual elements or individual offers in that set are linear. We can easily derive Eq.\ref{22} by taking any other set like this for another commodity. Let the offer set for other commodity is $ D_{2} $= {(1 kg, @ \rupee 55),(500g, @\rupee 20),(250g, @\rupee12)}. Then for each ordered pair of the set $ D_{1}\times D_{2} $=[\{(1 kg, @ \rupee 53.50)(1 kg, @ \rupee 55)\},\{(500g, @\rupee 27)(500g, @\rupee 20)\},\{(250g, @\rupee13.50)(250g, @\rupee12)\}] Equation \ref{22} can be calculated, and those will be the price response functions of the two commodities under each pair of the domain $ D_{1}\times D_{2} $. This will increase the predictability and the responses due to price, number of slabs, minimum order constraints, consumer motive, and budget. 
\paragraph{Non-negativity} The demand function $ x_{i}(p_{i}\mid\mu^{k},x_{i,min}^{k},m,p_{j},x_{j,min}	^{k};\forall j,k=1,...,N) $ is greater than or equal to zero for all prices i.e. $  x_{i}(p_{i})\geq 0 \forall p_{i}$.
\paragraph{Downward Sloping} There is a negative relationship exists between $  x_{i}(p_{i}) \&  p_{i}$ and its has been shown in Fig. \ref{figure39} and the equation of negatively sloped curve is $ x'(p_{i,min}^{k})= -[\dfrac{m\mu}{p^{2}}-\dfrac{p_{j}^{k}x_{j,min}^{k}\mu}{p^{2}} ]$ for $ [\dfrac{m\mu}{p^{2}}-\dfrac{p_{j}^{k}x_{j,min}^{k}\mu}{p^{2}} ]>0 $.
\paragraph{Continuity} It will have no gaps or jumps. The price response function proposed here has a small number of discrete steps because these are for only one customer. As the number of customers increases, the price response function will approach a more continuous function. While price-response functions are not truly continuous because prices of the discrete nature of customer demand and minimum order constraints and price can only occur at discrete intervals, they are often treated as continuous for mathematical convenience.
\paragraph{Hazard Rate} Hazard Rate of a price response function at a particular price is equal to minus one times the slope of the price response function at that price divided by the demand; i.e., $ h(p)= - \dfrac{x'(p)}{x(p)} $. This means the hazard rate is positive because the slope is negative. And the hazard rate is given by $ h(p)= -\dfrac{\mu(-m+p_{j}^{k}x_{j,min}^{k})}{(-m\mu+p_{i}^{k}x_{i,min}^{k}(\mu-1)+p_{j}^{k}x_{j,min}^{k}\mu)^{2}}$.
\paragraph{Price Elasticity}
The most common measure of the sensitivity of demand to price is price elasticity, defined as the ratio of the percentage change in demand to the percentage change in price. Formally, it can be written as $ \epsilon (p_{1},p_{2})=- \dfrac{100\{[x(p_{2})-x(p_{1})]/x(p_{1})\}}{100\{(p_{2}-p_{1})/p_{1}\}}$ where $  \epsilon (p_{1},p_{2}) $ is the elasticity of a price change from $ p_{1} $ to $ p_{2} $. However, the point elasticity at  $ p $ is written as $ \epsilon (p)=-\dfrac{x'(p_{1})p_{1}}{x(p_{1})} $.  In other words, the point elasticity equals $ 1 $ times the slope of the demand curve times the price divided by demand. And the Price Elasticity is given by $ \epsilon (p)= -\dfrac{p_{i}^{k}\mu(-m+p_{j}^{k}x_{j,min}^{k})}{(-m\mu+p_{i}^{k}x_{i,min}^{k}(\mu-1)+p_{j}^{k}x_{j,min}^{k}\mu)^{2}}$.
\paragraph{Price Response and Willingness to Pay} The willingness-to-pay approach assumes that each potential customer has a maximum willingness to pay (sometimes called a reservation price) for a product or service. A customer will purchase if and only if the price is less than her maximum willingness to pay. It can derive the willingness-to-pay distribution from the price-response function using\footnote{Assuming that the price-response function is finite.}
$ w(p) = - \dfrac{x'(p)}{x(0)} $. The demand at price $ p=0 $ is infinite and the limit value is also infinite i.e. $ \lim_{p_{i}^{k}\to\ 0} x(p_{i}^{k})= \infty \textit{sign} (m\mu-p_{j}^{k}x_{j,min}^{k}\mu)$. This means if $  \textit{sign} (m\mu-p_{j}^{k}x_{j,min}^{k}\mu)$ becomes zero, then it will be undefined. Otherwise, it will be positive infinity for positive signs and negative infinity for negative signs. Therefore the willingness to pay at $ p_{i}^{k}=0$ is $ 0 $. The willingness to pay exists at any positive limit value. Here, I have considered three possible cases. \\
(i) At $ p_{i}^{k}=0.01 $ the willingness to pay is $ \dfrac{\frac{m\mu}{p^{2}}-\frac{p_{j}^{k}x_{j,min}^{k}\mu}{p^{2}}}{10.0m\mu-10.0p_{j}^{k}x_{j,min}^{k}\mu-1.0x_{i,min}^{k}x_{j,min}^{k}+1.0x_{i,min}^{k}} $ \\
(ii) At $ p_{i}^{k}=0.001 $ the willingness to pay is $ \dfrac{\frac{m\mu}{p^{2}}-\frac{p_{j}^{k}x_{j,min}^{k}\mu}{p^{2}}}{100.0m\mu-100.0p_{j}^{k}x_{j,min}^{k}\mu-1.0x_{i,min}^{k}x_{j,min}^{k}+1.0x_{i,min}^{k}} $ \\
(iii) At $ p_{i}^{k}=0.001 $ the willingness to pay is $ \dfrac{\frac{m\mu}{p^{2}}-\frac{p_{j}^{k}x_{j,min}^{k}\mu}{p^{2}}}{100.0m\mu-100.0p_{j}^{k}x_{j,min}^{k}\mu-1.0x_{i,min}^{k}x_{j,min}^{k}+1.0x_{i,min}^{k}} $ \\
Therefore, it is clear that the willingness to pay is an increasing function with respect to price, given the other factors affecting the willingness to pay. Given the other factors, the willingness to pay can be increased by increasing the prices. This is so the customer has selected the particular slab and wants to stay put.  Once the preferred slab and the degree of preference of the product in question are known, the pricing will be the revenue maximizer.
\paragraph{Purchase probability.}I assume no offers in each slab have identical characteristics (the combination of price-quantity and conditional purchase probability). Given that position $ k $ is within a customer attention span, her purchase probability of the product at slab $ k$ is
\begin{equation*}
	\prod_{j=1}^{k-1}(1-\lambda_{j}).\lambda_{k}
\end{equation*}
where, $ \lambda_{k} $ is the probability that the customer will select a slab $ k $ given the prices and other factors affecting the demand. In other words, the customer purchases a product at the slab $ k$ if it is satisfactory while all products displayed earlier slabs are not. $ (1-\lambda_{k}) $ is the effect that a product at slab $ j  $ exerts on the products displayed later. This is the probability that the product at the $ j^{th} $ slab has not been preferred over $ k $. 
\section{Revenue Function}
The revenue function depends on the parameters viz. membership function, minimum order constraints of the commodity and the related commodities, prices	af of the commodity and the related commodities, and the choice of slabs by the customer. Moreover, consumers prefer to select the convex domain over the noon convex domain, which can be represented by the probability elements attached to it.  In this case, the revenue can be maximized by changing other factors, keeping the commodity's price exogenous. The platform controls the demand function using a choice-set policy. The platform's goal is to choose prices $ p_{i} $ to maximize the expected revenue from a potential consumer. From the earlier introduction, we see that the total expected revenue $ E[x_{i}(p_{i})] = R(x_{i})$  from a consumer with an individual demand function for each slab and with an attention span of $ y $ can be expressed as
\begin{equation}\label{23}
	\mathcal{R}(k, y,x_{i})\triangleq \sum_{k=1}^{\min(y,N)}\left[\prod_{j=1}^{k-1}(1-\lambda_{j})\right]\lambda_{k}.x_{i}(p_{i}^{k}) .p_{i}^{k} 
\end{equation}

\paragraph{Remarks 6} The Eq. \ref{23} is the expected revenue function and depends on the random choice of the slab of a given customized commodity offer. The predicted revenue function is affected by the addition and removal of slabs. This has been explained by explaining the creation of a non-convex domain to earn more revenue and push the customer to buy more, given the price in each slab.
\begin{proposition}
The expected revenue function is affected by the addition and removal of slabs. The higher the number of slabs, the higher the convexity, and the lower the expected revenue. On the other hand, the lower the number of slabs, the higher the non-convexity and the higher the expected revenue.
\end{proposition}
\begin{proof}
	The proof is in the appendix.
\end{proof}
\paragraph{Remarks 7} From Proposition 2, the expected revenue depends on the slabs' choice. Because the higher number of slabs will be supported by the lower prices, and on the other hand,  if the customer selects more elevated slabs for the lower prices, then the factor $ \left[\prod_{j=1}^{k-1}(1-\lambda_{j})\right]\lambda_{k} $ will be lower to get the expected value. This will again reduce the expected revenue. To maximize the expected revenue, there should be more than one slab to create the non-convex domain. As a result, the consumer will buy more quantity. On the other hand, if the slabs are increased, the customer will tend to buy less at a lower price. The slabs' choice depends on the willingness to pay and the intention to buy more.
This section generalizes the price-response function as defined in Eq.(16) with respect to different risk behavior.
\section{Conclusions and the Managerial Implications}
This article suggests a straightforward mechanism for slab pricing on an e-commerce platform. The article's main idea is that customers will be considered homogeneous if the quantity is provided in a compact set, which is closed and bounded. However, diverse customers can be produced if the same quantity is provided by making the domain non-convex. The study above indicates that income increases with increasing non-convexity. \\
\indent This means that in the case of any two commodities, as explained here, the revenue will be higher in terms of higher quantity sold. This is the case where \textit{Non-Convex Domain Case with Non-Linear $ X_{1} $ Price and Non-Linear $ X_{2} $ Price}; the less revenue will be earned in the case  \textit{Convex Domain Case with Linear $ X_{1} $ Price and Non-Linear $ X_{2} $ Price}, and finally the case \textit{Convex Domain Case with Linear $ X_{1} $ Price and Linear $ X_{2} $ Price}.\\
\indent The platform must choose the offering domain, figure out how many slabs to offer, and then calculate the equilibrium price for that domain to maximize income. The convex domain scenario will be more elevated, and the degree of nonlinearity will be lower the more slabs there are. Conversely, the greater the flexibility to apply non-linear prices, the fewer slabs there are. The revenue will drop as the number of slabs rises. Therefore, this article explains that choosing two slabs is the best option. However, the predicted revenue will drop if there is a more than two-slab increase. Consequently, the study suggests a novel method for figuring out non-linear prices by choosing the right amount of slabs, prices, minimum order restrictions, etc. A marketing manager has to remember that the least amount of money needed for matching must be spent when choosing the domain and spending to optimize revenue. If not, the offer would still be available, but the customer, assuming the platform employs the same method for choosing the domain and pricing, would go to another one where the matching price is the lowest.
\section{Limitation} 
The present article proposes a model of creating slabs by setting non-linear prices in each slab using a proposed elegant model. The present article does not consider the $ n $-commodity case. Therefore, in future research may be developed for $ n $ commodities. Conflict of Interest: we certify that we have no affiliations with or involvement in any organization or entity with any financial interest (such as honorarium, educational grants, participation in speakers bureaus, membership, employment, consultancies, stock ownership, or other equity interest; and expert testimony or patent-licensing arrangements), or non-financial interest (such as personal or professional relationships, affiliations, knowledge or beliefs) in the subject matter or materials discussed in this manuscript.\\
\textbf{Conflict of Interest}\\
I certify that we have no affiliations with or involvement in any organization or entity with any financial interest (such as honorarium, educational grants, participation in speakers Bureaus, membership, employment, consultancies, stock ownership, or other equity interest, and expert testimony or patent-licensing arrangements), or non-financial interest (such as personal or professional relationships, affiliations, knowledge, or beliefs) in the subject matter or materials discussed in this manuscript. \\
\textbf{Data Availability Statements}\\
I do not analyze or generate datasets because our work proceeds with a theoretical and mathematical approach. \\
\textbf{Appendix}\\
The proofs are available on demand. To access the proofs, you are to request the author via email at  \textit{E-mail:}\href{mailto:dipankar3das@gmail.com}{dipankar3das@gmail.com}   .
\bibliographystyle{apalike}
\bibliography{myreferences}
\end{document}